\documentclass[11pt,twoside]{article}
\usepackage{asp2010}

\resetcounters

\begin{document}

\title{Iron Depletion into Dust Grains in Galactic Planetary Nebulae}
\author{G. Delgado-Inglada and M. Rodr\'iguez}
\affil{Instituto Nacional de Astrof\'isica, \'Optica y Electr\'onica (INAOE), 
Apdo Postal 51 y 216, 72000 Puebla, Pue. Mexico}

\begin{abstract}
We present preliminary results of an analysis of the iron depletion factor into 
dust grains for a sample of 20 planetary nebulae (PNe) from the Galactic bulge. 
We compare these results with the ones we obtained in a prior analysis of 28 
Galactic disk PNe and 8 Galactic H II regions. 
We derive high depletion factors in all the objects, suggesting that more than 80\% 
of their iron atoms are condensed into dust grains. The range of iron depletions in 
the sample PNe covers about two orders of magnitude, and we explore here if the 
differences are related to the PN morphology. However, we do not find any significant 
correlation.
\end{abstract}

\section{Introduction}
Asymptotic giant branch (AGB) stars suffer mass loss episodes during the late 
stages of their evolution. The combination of high densities with relatively low 
temperatures makes the atmospheres of these stars favourable sites for grain 
formation, and in fact they are considered one of the most efficient sources of dust 
in the Galaxy \citep[see, e.g.,][]{Whittet_03}.
Planetary nebulae (PNe) are created when AGB stars reach temperatures high enough to 
ionize the material previously ejected. 
Several authors have studied the dust present in PNe, but it is not clear how much 
dust they have and whether it is destroyed or modified during their lifetimes 
\citep{Pottasch_84, Lenzuni_89, Stasinska_99}.

We study the dust in PNe through the analysis of the iron depletion factor, 
the ratio between the expected abundance of iron and the one measured in the 
gas phase. 
The abundances of refractory elements like iron have been calculated from ultraviolet 
absorption lines in several interstellar clouds toward different sight lines, and the 
low values obtained, compared to 
the solar ones, are generally interpreted as due to depletion into dust grains \citep[see, e.g.,][]{Morton_74}. 
Iron is mostly condensed into grains and has a 
relatively high cosmic abundance. Those two facts together make iron an important 
contributor to the mass of refractory grains \citep{Sofia_94}, and hence the iron 
depletion factor is likely to reflect the abundance of refractory elements in dust 
grains. Besides, in ionized nebulae iron is the refractory element with the strongest 
lines in the optical range of the spectrum.

In a previous work \citep{Delgado-Inglada_09}, we performed a
homogeneous analysis of the iron abundance in a sample of 28 Galactic
disk PNe and 8 Galactic H II regions. We obtained very low iron
abundances in all the objects, implying that more than 90\% of their
total iron abundance is condensed into dust grains. This suggests that
iron depletes very efficiently in AGB stars and molecular clouds,
whereas refractory dust is barely destroyed in the ionized gas of PNe
and H II regions. Here, we extend our analysis to include 20 Galactic
bulge PNe (observed by \citealt{Wang_07}), and study the dependence of
the results on PN morphology.

\section{The analysis}
In H II regions and low ionization PNe, Fe$^{++}$ and Fe$^{+3}$ are the dominant  
ionization stages of iron. Due to the faintness of [Fe IV] lines, the gaseous iron 
abundance is usually calculated from the Fe$^{++}$ abundance and an ionization correction 
factor (ICF) derived from photoionization models. However, for the handful of objects 
with measurements of [Fe III] and [Fe IV] lines, a discrepancy has been found between 
the abundance calculated with the aforementioned method and the one found by adding 
the abundances of Fe$^{++}$ and Fe$^{+3}$. 
\citet{Rodriguez_05} studied this discrepancy and derived three correction schemes 
that take into account what changes in all the atomic data involved in the calculations 
would solve the discrepancy: 1) a decrease in the collision strengths for Fe$^{+3}$ 
by factors of $\sim$2--3, 2) an increase in the collision strengths for Fe$^{++}$ by 
factors of $\sim$2--3, or 3) an increase in the total recombination coefficient 
or the rate of the charge-exchange reaction with H$^0$ for Fe$^{+3}$ by a factor of 
$\sim$10. 
Since these three possibilities are equally plausible, and since the discrepancy 
could be due to some combination of them, the real value of the iron abundance will 
be intermediate between the extreme values derived with the three correction schemes.

As in our previous work, we select PNe with a relatively low degree of ionization 
($I(\mbox{He II}\,\lambda4686)/I(\mbox{H}\beta)$ $\lesssim0.3$, 
see \citealt{Delgado-Inglada_09}), and with moderate electron density (below 
25\,000 cm$^{-3}$), since high densities can be associated with large density 
gradients that introduce large uncertainties in the calculations. Besides, the objects 
have spectra with all the lines we need to calculate the physical conditions and the 
abundances of Fe$^{++}$, O$^{+}$, and O$^{++}$.
We calculate a mean electron density and two electron temperatures (using the usual 
[N II] and [O III] diagnostic line ratios) for the low and high ionization regions; 
and with them we derive the O$^{+}$, O$^{++}$, and Fe$^{++}$ abundances. 
To derive the total oxygen abundances we use the ICF from 
\citet{KB_94} and for iron the three ionization schemes of \citet{Rodriguez_05}. 
See \citet{Delgado-Inglada_09} for more details on the procedure and on the atomic 
data we use. The uncertainties in all the quantities were estimated via Montecarlo 
simulations (Delgado-Inglada \& Rodr\'iguez, in prep.).

\section{Results}
\begin{figure}
\epsscale{1.0}
\plotone{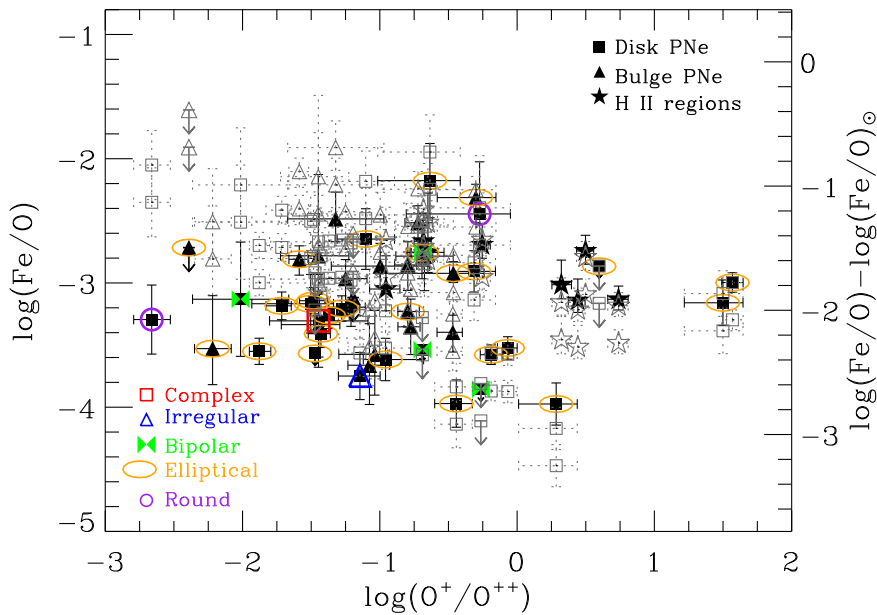}
\caption{{\small Values of Fe/O (left axis) and the depletion factors 
for Fe/O ($[\mbox{Fe/O}]= \log(\mbox{Fe/O}) - \log(\mbox{Fe/O})_{\odot}$, right 
axis) as a function of the degree of ionization. Three values of Fe/O are shown for 
each object, derived using the three correction schemes of \citet{Rodriguez_05}. 
The PNe are classified according to their origin (disk or bulge) and their morphology. 
See the text for more explanations.
\label{fig1}}}
\end{figure}

Figure \ref{fig1} shows the values of the Fe/O abundance ratio (left axis), and 
the depletion factors (right axis) for Fe/O: 
$[\mbox{Fe/O}]= \log(\mbox{Fe/O}) - \log(\mbox{Fe/O})_{\odot}$. We use the solar 
abundance from \citealt{Lodders_03}, $\log(\mbox{Fe/O})_{\odot} = -1.22\pm0.06$, 
as the expected total abundance.
Filled symbols in this figure indicate the values derived using the correction scheme 
defined by point (3) above, while empty symbols represent the values obtained with 
the other two correction schemes. The real values are expected to be in the range 
defined by these three values. We find that both PNe and H II regions have 
consistently high depletions factors, with more than 80\% of their iron atoms 
condensed into dust grains. 
The differences between the Fe/O values derived with the three correction schemes 
depend on the degree of ionization of the objects: the higher the degree of ionization, 
the greater the difference in Fe/O values. Hence, the iron abundance is better 
constrained for the objects with $\log$(O$^{+}$/O$^{++}$) $\gtrsim-1.0$, where 
the differences between the correction schemes are $\leq$0.5 dex. No matter which 
correction scheme we consider, the range of depletions is wide in this region, with a 
difference in the iron abundance of a factor of $\sim$100 between the objects with 
the lowest and the highest value.
Although these variations should be reflecting real differences between the objects, 
we do not find any clear correlation between the iron abundances and parameters related 
to the nebular age (such as the electron density or the surface brightness of the PN) 
or with the dust chemistry (Delgado-Inglada \& Rodr\'iguez, in prep.). 
The first result suggests that no significant destruction of dust grains is taking 
place in these objects, and the second one argues for a similar efficiency of iron 
depletion in C-rich and O-rich environments. 

Here, we explore the relation between the iron abundances of the sample PNe and their 
morphological types. There is some observational evidence that the morphological type 
of a PN is related to the progenitor mass. Symmetric PNe (round or elliptical nebulae) 
might descend from low mass progenitors, and asymmetric PNe (bipolar or more complex 
objects) might have the most massive progenitor stars 
\citep[see, e.g.,][]{Corradi_95, Stanghellini_06}. Asymmetric PNe have also been 
associated with binary systems \citep[see, e.g.,][]{Corradi_95, Soker_98, deMarco_09}, 
and with stellar rotation and/or magnetic fields \citep[see, e.g.,][]{GarciaSegura_99}. 
Although the reasons behind the shaping of PNe are still matter of debate, we study 
here if there is a correlation between the morphologies and the iron depletion factors.
Figure \ref{fig1} shows the PNe morphologies given by the Planetary Nebula Image Catalog 
of Bruce Balick (PNIC\footnote{http://www.astro.washington.edu/users/balick/PNIC/}) 
and by \citet{Stanghellini_02} for the 34 objects with available images. 
Two of them are classified as round, 26 as elliptical, four as bipolar, one as 
irregular, and one more as complex. This figure shows that there is no obvious trend 
relating the morphological type and the amount of iron condensed in dust grains.
For example, elliptical and bipolar PNe are distributed in the whole 
range of depletions. The three correction schemes used in the abundance determination 
give consistent results on this issue.

\section{Conclusions}

We derive the iron abundance in a sample of 20 Galactic bulge PNe,
and compare the results with the ones previously obtained for 28
Galactic disk PNe and 8 Galactic H II regions. We find high depletion
factors in all the objects: less than 20\% of their expected total
number of iron atoms are measured in the gas phase. This result
suggests that iron depletion into dust grains is an efficient process
in AGB stars, whereas dust destruction is not very efficient in the
subsequent PN phase. Although the range of depletions in the sample
PNe is wide, covering around two orders of magnitude, we do not find
any correlation between the iron depletion factors and the morphology
of the PNe. Further details of the analysis will we presented in
Delgado-Inglada \& Rodr\'iguez (in prep.).

\acknowledgements We acknowledge support from Mexican CONACYT project 50359-F.

\bibliography{Delgado-Inglada}

\end{document}